# Sense current dependent coercivity and magnetization relaxation in Gd-Fe-Co Hall bar

Ramesh Chandra Bhatt,[1,2] Chun-Ming Liao,[1,2] Lin-Xiu Ye,[1,2] Ngo Trong Hai,[3] Jong-Ching Wu,[3] and Te-ho Wu,[1,2,a]

[1]Graduate School of Materials Science, National Yunlin University of Science and Technology, Douliu, Yunlin 640 Taiwan ROC

[2]Taiwan SPIN Research Center, National Yunlin University of Science and Technology, Douliu, Yunlin 640, Taiwan ROC

[3]Department of Physics, National Changhua University of Education, Changhua 500, Taiwan ROC

[a]**Author to whom correspondence should be addressed:** wuth@yuntech.edu.tw

**ABSTRACT**

The understanding of the characteristics of a magnetic layer in a different environment is crucial for any spintronics application. Before practical applications, thorough scrutiny of such devices is compulsory. Here we study such a potential Hall device of MgO-capped Hf/GdFeCo bilayer (FeCo-rich) for magnetization relaxation around nucleation fields at different voltage probe line widths and dc sensing currents. The device is characterized by anomalous Hall measurements in transverse and longitudinal Hall geometries for two different probe widths A (5 µm) and B (1 µm). The coercivities of the Hall loops ($\rho_{xy}$-$H$ and $R_{xx}$-$H$) drop with increasing the sense current for both the probes. For probe B, the sharp and large drop in coercivity ($\rho_{xy}$-$H$ loops) at comparatively lower sensing currents is observed, which is attributed to the negligible current shunting and presence of pinning site at B caused by the patterning process. The average domain wall velocities at various sensing currents for probe B are found to be smaller than probe A, from the transverse and longitudinal Hall geometry magnetization relaxation measurements, which agrees with pinning sites and Joule heating effect at probe B. The notch position in the pattern and the longitudinal Hall resistance curve peak shape suggest the domain wall propagation direction from probe B to probe A in the current channel. This study highlights the domain wall propagation at different nucleation fields, sensing currents, and the Hall probe aspect ratios.

**INTRODUCTION**

Most of the technologies we use in daily life in communication, display, transport, health, banking, social media, electronic gadgets, etc., are based on magnetic sensors and memories[1]. Interestingly, the majority of these magnetic materials belong to the ferromagnetic family of magnetism. However, a special class of ferrimagnets that is, the rare earth (RE)-transition metal (TM) alloys, which were known long before for magneto-optical recording[2,3], have recently attracted much attention due to their ultrafast magnetization dynamics[4,5], easy manipulation of magnetization by the electrical current[6–10], laser[11,12], etc., leading to many applications such as, Terahertz emitter[13], spin-torque nano-oscillator[14], magnetic random access memory, neuromorphic computing[15], etc. The RE (Tb, Gd, etc.)-TM (Fe and Co) alloys have f and d-electrons in the RE and TM sublattices, respectively. In general, a thin film of these materials shows strong perpendicular magnetic anisotropy (PMA) with an anisotropy energy density of about $10^6$ erg/cm$^3$[16,17]. The two sublattices strongly exchange by a negative exchange interaction, and the net magnetization ($M_S$) of the alloy can be varied from zero to maximum by adjusting the RE/TM content ratio from compensation to RE/TM dominant composition, respectively[18–20]. At compensation, with vanishing $M_S$, their magnetization dynamics is as fast as antiferromagnet, yet can be sensed by the Hall measurements through the TM sublattices' d-electrons[5,10,21–25]. For spintronics application purposes a basic device consists of the Hall bar pattern, which can be used for anomalous Hall effect (AHE), spin Hall switching, spin torques, domain wall velocity, etc., important characteristics[6,26,27]. In the Hall bar device, the effect of sense current on anomalous Hall effect resistivity, coercivity, domain wall propagation, are key factors for the memory and other spintronics applications[28,29]. Furthermore, these characteristics also depend on the shape of the pattern (aspect ratio)[28–30]. In our previous studies, we have investigated the $H_C$, AHE resistivity ($\rho_{AHE}$), and magnetization relaxation behavior as a function of sense current and Hall bar aspect ratios in the TbFeCo[28,29]. In the present work, we investigate GdFeCo thin film through AHE measurements to find the dependence of $H_C$, $\rho_{AHE}$, and the time-evolution of magnetization on the sense current and Hall bar aspect ratio. This work deals with controlled propagation of domain walls therefore the domain wall velocity is much slower in comparison to other ways of switching the magnetization[10,29].

**MATERIALS AND METHODS**

The MgO capped Hf/GdFeCo bilayer film, the schematic as shown in the left side image of Fig. 1(a), was deposited on thermally oxidized Si-substrates using a high vacuum dc/rf magnetron sputtering (base pressure 1.5 x 10$^{-7}$ Torr). The GdFeCo was co-sputtered from the Gd and Fe$_{80}$Co$_{20}$ targets at 70 Watt and 200 Watt dc sputtering power, respectively. The substrate holder was spinning and rotating simultaneously during the deposition. The magnetic properties of the deposited film were measured using an Alternating-Gradient Magnetometer (PMC AGM) at room temperature which showed perpendicular magnetic anisotropy in the alloy film. The Hall bar device for the anomalous Hall measurements was fabricated by electron-beam lithography and Ar-ion milling. The optical image of the Hall bar is shown on the right side of Fig. 1(a), with various lengths and widths of the Hall bar. The current channel has a notch of 2 µm width in the Hall bar. Two voltage probe lines have different widths of 5 µm and 1 µm, calling these probes 'A' and 'B', respectively. The anomalous Hall resistivity ($\rho_{AHE}$) measurements at probe 'A' and 'B' were carried out using dc sensing current in an electromagnet-equipped four-probe station at room temperature (300 K). The longitudinal resistance $R_{xx}$ was measured between the Probe 'A' and 'B' in the current direction. The current was limited to 3 mA to avoid the circuit breakdown and/or the rectangular shape deformation of the AHE curve.

**RESULTS AND DISCUSSION**

The anomalous Hall resistivity is measured at the two probes 'A' and 'B' in the device for the different dc sensing currents (10 µA to 2.5 mA). The $\rho_{AHE}$ loops at different sensing currents are shown in Fig. 1(b). The magnetic field offset ~ 10 Oe can be seen in all the loops which correspond to the instrument remanence and not from the device. The shape of the loops for the probe 'A' remains rectangular for the whole current range (i.e., 10 µA – 2.5 mA). However, for probe 'B' loops, the rectangular shape starts deforming after 1 mA. No effect of current on the magnitude of $\rho_{AHE}$ is seen for the two probe lines as the $\rho_{AHE}$ (5.3 µΩ.cm and 4.9 µΩ.cm for A and B, respectively) remains nearly unchanged on increasing the current, except for probe B at 2.5 mA, for which the $\rho_{AHE}$ drops from 4.9 µΩ.cm to 4.6 µΩ.cm. However, coercivity changes significantly which is shown in Fig. 1(c). At large currents, over 1.5 mA, the coercivity starts suddenly dropping for the probe 'A'. However, in the case of probe 'B', the sharp drop in coercivity is witnessed over 500 µA. At 2.5 mA, the coercivity for probe 'A' is 29.0 Oe, on the other hand, for probe 'B' it is much lower 19.7 Oe. At 50 µA, a small peak in the $H_C$ plot is observed for both the probes. The aspect ratio of the probes is defined as the ratio of voltage pickup line width to the current channel width. In the present case, the current channel width is the same for the two probes. Therefore, the aspect ratios for probe 'A' and 'B' in the Hall bar are 1 (5 µm/5 µm) and 0.2 (1 µm/5 µm), respectively. Therefore, more current shunting is expected for probe A than probe B. Our previous study on MgO-capped TbFeCo (Tb-rich)/Ta bilayer Hall bar device shows a drop in coercivity as well as in anomalous Hall resistivity for an increase in the dc sensing current, which has been explained on behalf of Joule heating and spin-orbit torque (SOT) effects[28]. In addition, the difference in the coercivity and anomalous Hall resistivity behavior on mutually exchanging the current and voltage probes has been attributed to the different current shunting in the Hall bar for different aspect ratios. As the aspect ratio for probe B is much smaller than A, the current shunting effect is negligible for probe B. Therefore, we expect more Joule heating effect as well as current-induced SOT for probe B, which explains the large drop in the $H_C$. Further, it also explains the drop in $\rho_{AHE}$ at 2.5 mA for probe B. Moreover, $\rho_{AHE}$ is proportional to the perpendicular component of magnetization ($m_z$) which drops due to the effective Joule heating and current-induced SOT. In a typical $H_C$-T curve of a RE-TM sample, the coercivity diverges at compensation temperature ($T_{comp}$). Below $T_{comp}$, the sample shows RE-rich behavior, whereas, above $T_{comp}$, it shows TM-rich behavior. Therefore, in a TM-rich sample, which has $T_{comp}$ just below the room temperature, a sharp drop in coercivity and $1/M_S$ can be seen with current-induced Joule heating. The decreasing $1/M_S$ should reflect as an increasing $\rho_{AHE}$ with Joule heating. Therefore, for a RE-rich sample: $H_C \uparrow$, $M_S \downarrow$ ($\rho_{AHE} \downarrow$) on increasing the electrical current (Joule heating), and for a TM-rich sample: $H_C \downarrow$, $M_S \uparrow$ ($\rho_{AHE} \uparrow$) on increasing the current. However, another factor is common for both the cases, that is, $H_C \downarrow$, $M_S \downarrow$ ($\rho_{AHE} \downarrow$) on increasing the current due to the SOT effect.

In the present case, the $H_C$ drops with increasing current but not the $\rho_{AHE}$, which should also be affected as being proportional to $m_Z$. However, in our case, the current-induced spin current from the Hf-layer could also assist FeCo magnetization switching, resulting in a drop in the $H_C$. Both the Joule heating and SOT favor a drop in $H_C$ which is evident from the results but the unchanged $\rho_{AHE}$ is the combined effect of both the Joule heating and SOT.

We further examined the time-evolved anomalous Hall resistivity at 40 Oe nucleation field (including instrument offset field ~ 10 Oe) at the two different probes 'A' and 'B'. These experiments were carried out for the different sensing current ranging from 10 µA to 2 mA. A detailed explanation of the experimental procedure can be found in our previous work[29]. Figure 2(a) shows the change of $\rho_{AHE}$ from $m_Z$ down to $m_Z$ up state for 10 µA current, which takes time to saturate. For 50 µA, the nucleation starts at the same time, as of 10 µA, with spending some time on intermediate points yet saturating earlier than the 10 µA curve. For 100 µA, nucleation starts much later but takes lesser time on saturating to the switched state. There are some intermediate points before saturation. On reaching 500 µA, nucleation starts earlier than 100 µA and saturates through intermediate points. For 1 mA, nucleation starts earlier than preceding cases and it spans through many intermediate points before complete switching. On further increase in sense current (i.e., 1.5 mA and 2 mA), the nucleation, as well as the saturation, occurs earlier than the preceding currents.

For probe 'B', as shown in Fig. 2(b), for 10 µA, the switching does not take place within the time frame shown in the inset. However, for 50 µA, the switching takes place with multi intermediate states before saturation, which appears like ladder-steps in the curve. These ladder-steps possibly arise from the pinning centers created due to the magnetic or geometric defects in the Hall bar pattern. For 100 µA, the nucleation takes maximum time to start and the intermediate step becomes more distinct. On further increasing the current the intermediate steps seem disappeared and the nucleation as well switching takes place at a much earlier time. After 1 mA, the switching (or the domain wall propagation) is so fast that it is not possible to record it from the existing setup.

From Fig. 2, we can calculate the time taken from nucleation to the saturation state, which is shown in the inset of Fig. 3. Here, we have excluded the nucleation delay time in calculating the switching time. The switching time is equivalent to the average time taken by the domains crossing the probe A (or B) in the current channel is shown in Fig. 3. Therefore, one can get the average domain wall velocity ($v_{DW}$) at these probes for different sensing currents. The $v_{DW}$ at these two probes is shown in Fig. 3. The $v_{DW}$ increases with the current for both the probes. However, for probe 'B' the velocities are much smaller than 'A'. This difference is due to the presence of several intermediate magnetization states during the magnetization relaxation at probe 'B, which possibly due to the pinning sites. Moreover, we speculate that the current shunting at probe 'A' is significant to have a curl in the current, which generates a perpendicular magnetic field in the negative *z*-axis, assists the magnetization switching.

We observed asymmetric anisotropic magnetoresistance (AMR) in our previous work, therefore, here also we investigated the device for the longitudinal Hall resistance ($R_{xx}$) measurements between probe A and B. The asymmetric AMR loops at different sensing currents ranging from 10 µA to 3 mA are shown in Fig. 4(a). At large electric currents, the peak gets distorted and peak widths increase, which suggests the Joule heating effect causing a multi-domain effect in the device. We defined here the coercivity of these asymmetric AMR loops as half of the peak to peak field value. The loop's coercivity is shown in Fig. 4(b). The coercivity increases initially for small currents up to ~ 33 Oe at 500 µA and rapidly decreases afterward. We also measured the magnetization relaxation at different nucleation fields at a 10 µA sensing current, as shown in Fig. 5(a). The nucleation fields have been chosen from the curve where the nucleation starts. The nucleation field includes an instrument offset field (~ 10 Oe), therefore, seems larger than the coercivity. Before starting the measurement, the device is saturated to -$m_Z$ direction by applying a negative perpendicular field. Therefore, to switch the magnetization a positive nucleation field is given. At 30 Oe nucleation field, the $R_{xx}$ drops slightly but it does not reach the valley point till ~74 seconds, which is our maximum programmed time in the instrument. When we measure the 35 Oe nucleation field, the $R_{xx}$ drops in steps to the valley point and remains at that state. However, on setting the nucleation field to the 40 Oe in the next measurement, the $R_{xx}$ drops from initial resistance to the valley point quickly and then returns to the high resistance state afterward.

Further, for a fixed nucleation field the magnetization relaxation process at different sensing currents is carried out as shown in Fig. 5(b). At 10 µA, the resistance drops in steps to a minimum value and remains in that state. For 50 µA and 100 µA sensing currents, the resistance drops around ~ 50 sec in multiple steps to a minimum point. Similarly, for 500 µA, with an early drop in the resistance. For 1 mA, the resistance lowers to a minimum value for some time and return to the high resistance state. The same process occurs much faster when the sense current sets to 1.5 mA. The returning resistance state corresponds to the domain-crossing through the second probe or switching across the second probe[29]. The magnetization curve measured at 10 µA sensing current and 40 Oe nucleation field, is further analyzed for the average domain wall velocity in the current channel as shown in Fig. 6. In the curve, the saturation state corresponds to the state when the domain wall crosses both the voltage probes A and B. And the valley of the curve corresponds to the crossing of either probe A or B. When we compare the bifurcated curve with Fig. 2, we see that the left side of the curve shows many intermediate states as in the case of Fig. 2(b) for probe B. Therefore, the possible domain wall motion is opposite to the current direction and it starts from probe B to probe A. Consequently, the right-side curve represents the distance traveled from the left end of probe B to the left end of probe A (i.e., the sum of the distance of interprobe distance and probe A width). Therefore, the saturation state to the valley point on the left side of the curve represents probe A width (1 µm) and the right side of the curve represents the interprobe distance (30 µm) + probe B width (5 µm). Fig. 6(b) and (c) show such an analyzed curve and the corresponding average domain wall velocity. The domain wall velocity is much slower for probe B than the probe A.

This is because initially, the nucleation field takes time to overcome the pinning sites at probe B which are present there due to the geometry defects. Once crossing the probe B domain wall does not face such magnetic inhomogeneity and propagates much faster through the current channel.

## CONCLUSIONS

In summary, we have fabricated the Hall bar pattern of MgO capped Hf/GdFeCo bilayer and studied its magnetization relaxation behavior at different probe widths, sensing currents, and transverse and longitudinal Hall geometries. The contributions from current shunting, Joule heating, and current-induced SOT are responsible for coercivity response to different sense currents in RE or TM-rich films, which also manipulates magnetization relaxation in the device. The negligible current shunting and consequently more Joule heating and current-induced SOT at thinner probe provide an early drop in the coercivity. In addition to that, the pinning sites available due to the patterning process at probe B reveals intermediate states in the magnetization relaxation. The domain wall propagation is faster for probe A than probe B, which increases with increasing the sense current. The asymmetric distortion in the peak shape of longitudinal magnetization relaxation curve suggests domain wall propagation from the right (probe B) to the left (probe A) in the current channel with delayed propagation at probe B. The notch on the left side of the pattern might be supporting the domain wall propagation opposite to the current direction. The present study on GdFeCo Hall bar behavior at different circumstances is crucial for their practical use in spintronics applications such as a magnetic memory element.

## ACKNOWLEDGMENT

This work was supported by the Ministry of Science and Technology (MOST) Taiwan ROC (Grant Nos. MOST 107-2112-M-224-001-MY2 and MOST 109-2112-M-224 -001 -MY2) and the Feng-Tay foundation Taiwan ROC (Grant No. 150-F32-3).## REFERENCES

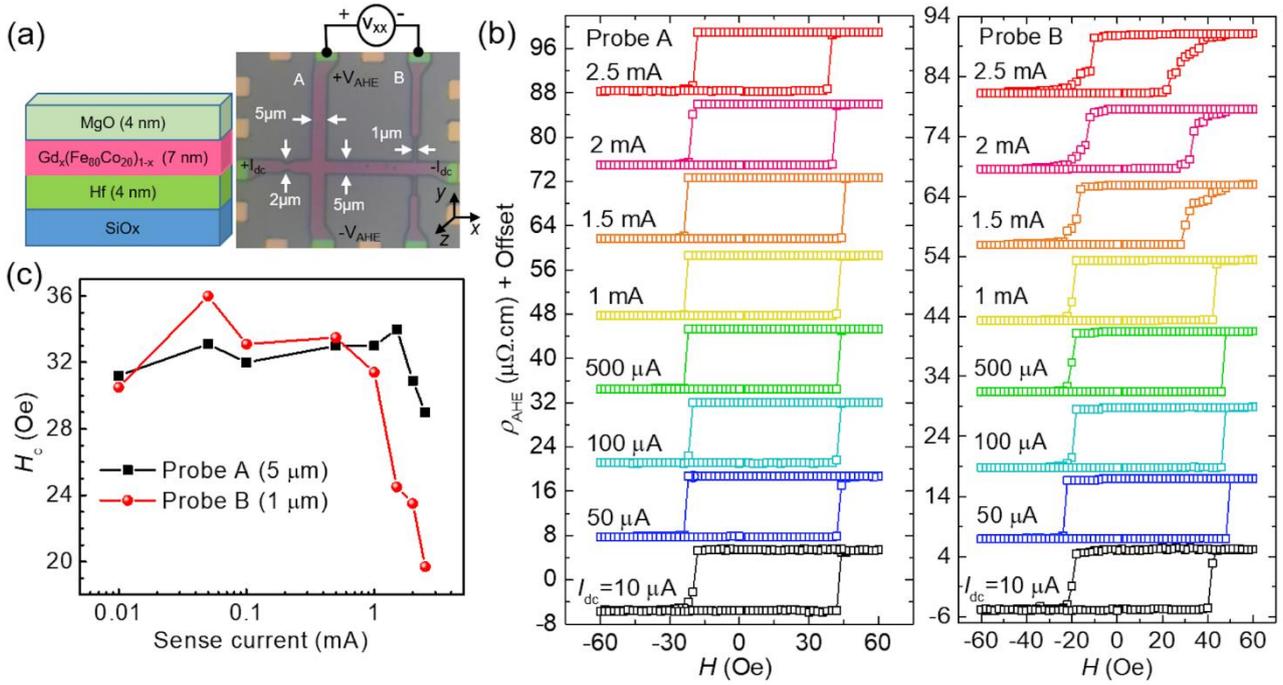

**Fig. 1** (a) Schematic of thin-film structure (left side) and optical image of Hall bar device (right side) indicating the various bar widths and connections for longitudinal ($V_{XX}$) and transverse ($V_{AHE}$) voltage measurements. The voltage pickup lines for the probe width of 5 μm and 1 μm are indexed as 'A' and 'B' in the optical image. (b) Anomalous Hall effect resistivity ($\rho_{AHE}$) at different sensing currents measured at two different probes 'A' and 'B'. In both the $\rho_{AHE}$-loop sets, the 10 μA loop is in the proper scale and the other loops are successively shifted vertically upwards. (c) $\rho_{AHE}$-loop coercivity at different sensing current for the two probe widths.

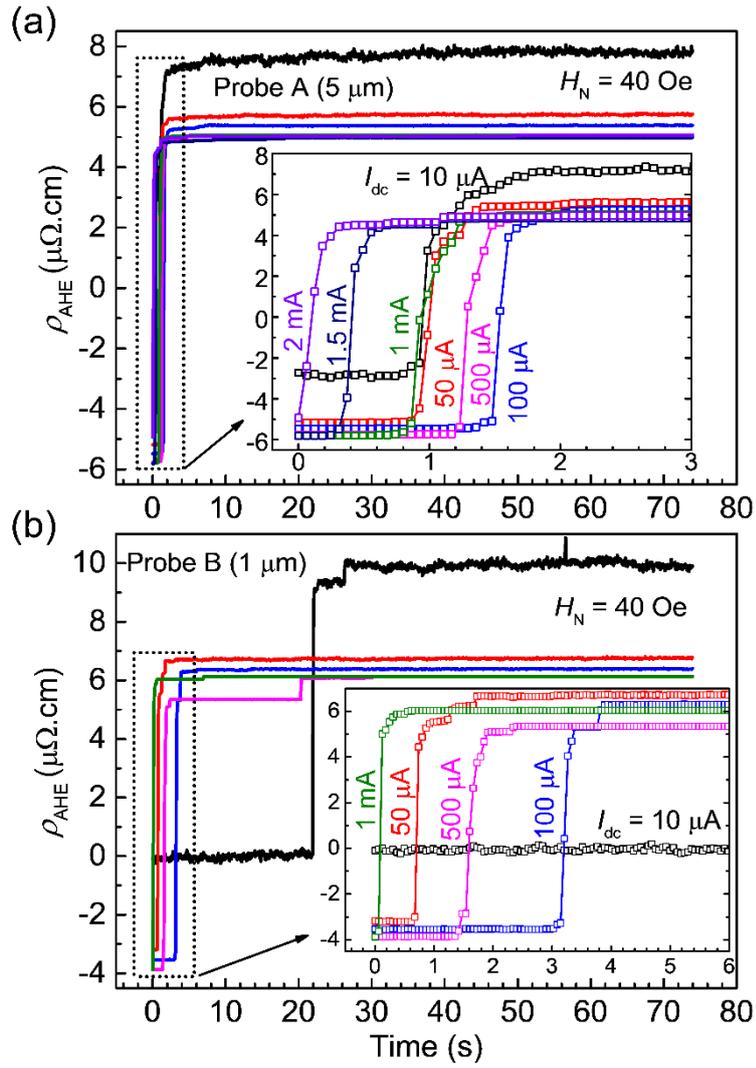

**Fig. 2** AHE resistivity time-evolution at constant perpendicular field for various sensing currents. AHE resistivity measured at (a) probe width of 5 μm and (b) probe width of 1 μm. The corresponding figure insets show the magnified view to resolve the magnetization switching at different sensing currents.

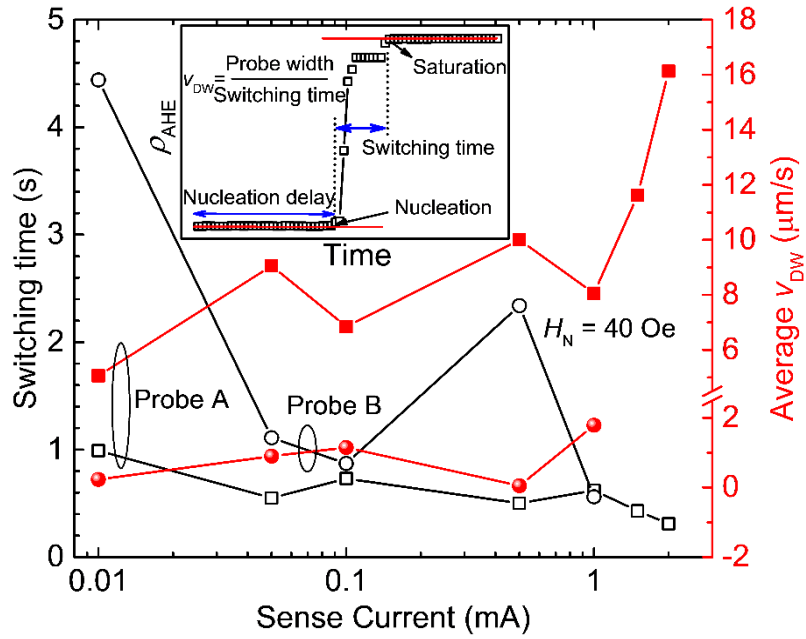

**Fig. 3** The domain wall propagation time characteristics at 40 Oe nucleation field. The inset shows the definitions of time delay in nucleation process, the switching time (i.e., the time taken from nucleation to the saturation of the magnetization), and the average domain wall velocity ($v_{DW}$). The switching time decreases with increasing sensing currents, results increase in the average $v_{DW}$ for the probes A and B.

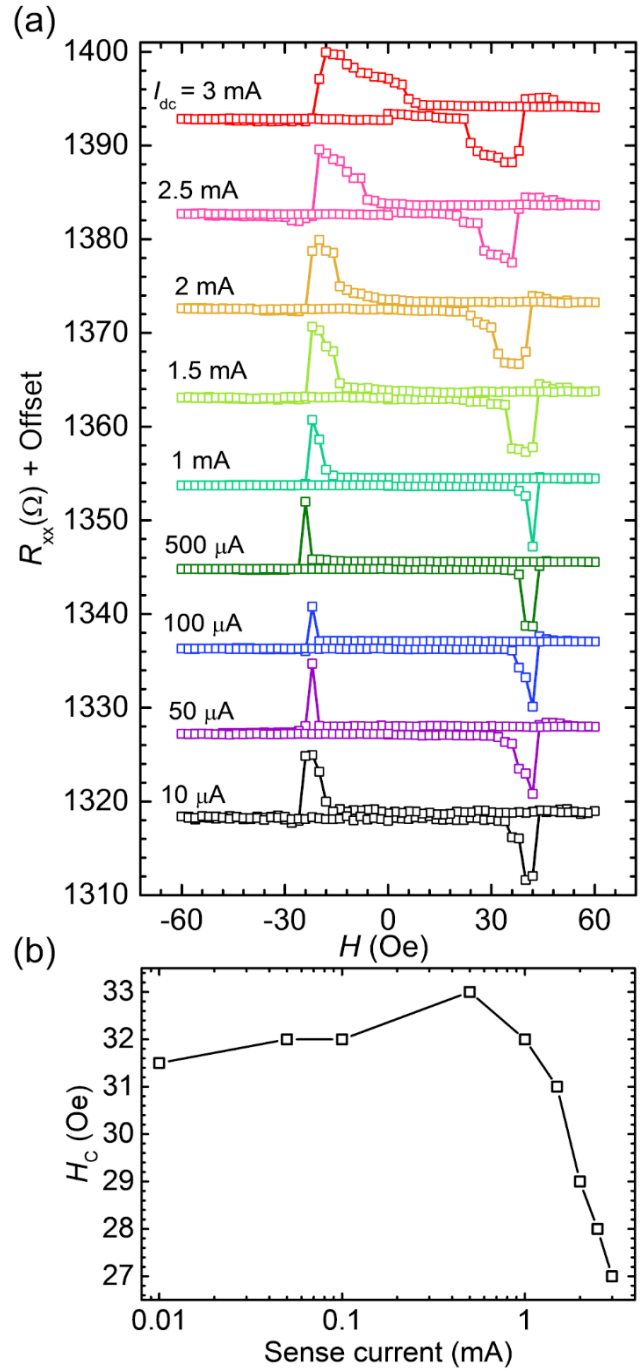

**Fig. 4** (a) Longitudinal resistance ($R_{xx}$) loops at various sensing currents measured between probe 'A' and 'B'. (b) Variation of coercivity with sensing currents obtained from the corresponding longitudinal resistance loops. Here, the coercivity is defined as the half of the peak to peak magnetic field values.

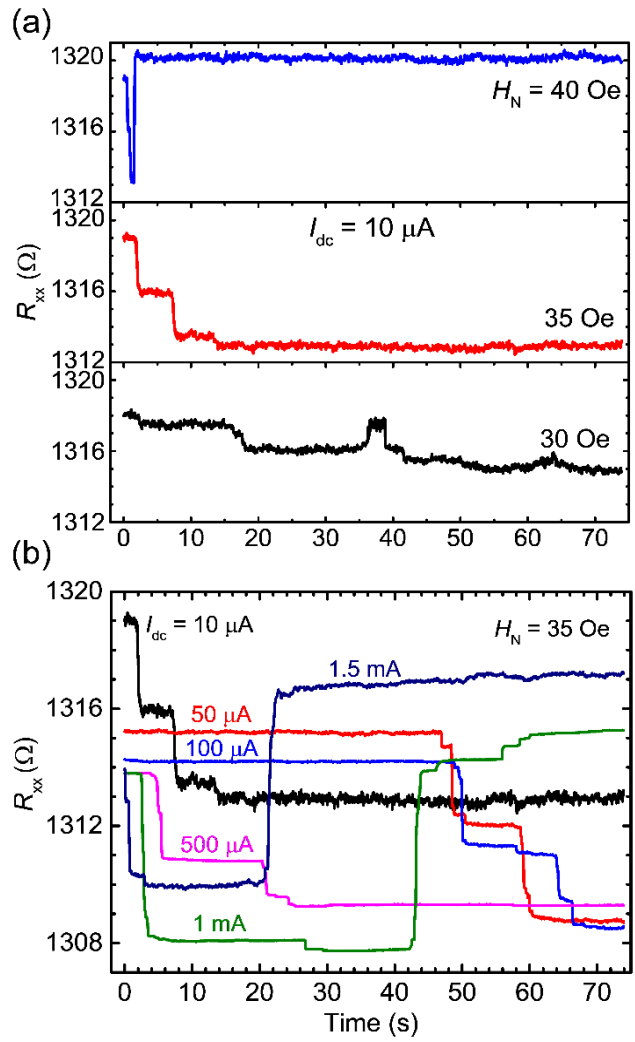

**Fig. 5** (a) Longitudinal resistance time-evolution for 10 µA dc sensing current at various nucleation fields. (b) Longitudinal resistance time-evolution for different sensing currents at 35 Oe nucleation field.

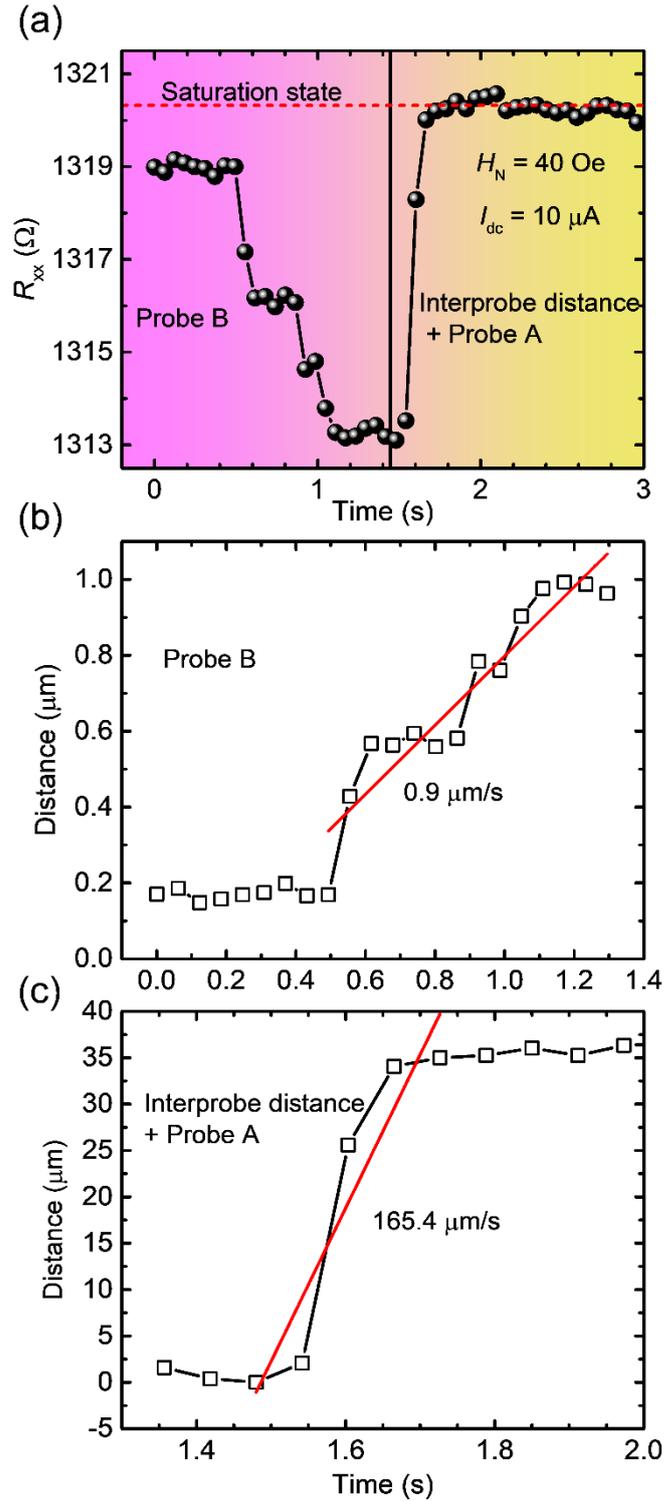

**Fig. 6** (a) A magnetization relaxation curve measured in Longitudinal Hall geometry at 40 Oe nucleation field (Fig. 5(a)). Bifurcating the curve and converting it to corresponding distance vs time plot: The distance traveled by domain wall on crossing (b) probe B (1 μm) and (c) interprobe distance + probe A (total 31 μm) in the current channel, respectively. The linear fit to the slope shows the average domain wall velocity corresponding to partitioned curves.